\documentclass[a4paper]{easychair}
\usepackage[verbose]{placeins}
\usepackage{doc}

\usepackage{color}

\title{Towards Truck Parking Lot Occupancy Estimation}

\author{
    Florian Ziegler
\and
    Maurice Freund
\and
    Andreas Rydzek
\and
   Thomas Liebig
}

\institute{
	Materna Information \& Communications SE, \\
  Dortmund, Germany\\
  \email{\{firstname.lastname\}@materna.de}
}

\authorrunning{Ziegler et al.}

\titlerunning{Towards Truck Parking Lot Occupancy Estimation}

\begin{document}

\maketitle

\begin{abstract}
  Occupied truck parking lots regularly cause hazardous situations. Estimation of current parking lot state could be utilized to provide drivers parking recommendations.
  In this work, we highlight based on a simulation scenario, how sparse observations, as obtained by a mobile application could be utilized to estimate parking lot occupancy. Our simulated results reveal that a detection of a filled parking lot could be possible with an error of less than half an hour, if the required data would be available.
\end{abstract}



%
%

\section{Introduction}
\label{sect:introduction}

In recent years, the increase in street-based cargo shipping in combination with European and national regulation of drivers working hours (e.g. \cite{union2006verordnung}) led to an increased demand for parking capacities.
Often, parking slots are occupied when a driver arrives, and she has to move further to check availability at the next parking lot or hazardous situations occur \cite{haque2017truck}.

The problem is crucial: In a representative study \cite{itp}\footnote{Study carried out in conjunction with the intelligent truck parking project funded by the German Federal Ministry of Transport and Digital Infrastructure (BMVI) program mFUND under grant number FKZ: 19F2037B.}, about 64 percent of truck drivers states they are regularly missing a free parking lot, and every fourth of the drivers has difficulties full-filling slot restrictions due to parking lot search.
A simple solution to the problem would be the collection of floating car data from the trucks or obtaining count values directly from the parking infrastructure, as performed in \cite{haque2017truck}. However, this requires the proliferation of multiple devices or access to telematics of the trucks.

Instead, the ITP project supposes the usage of a parking app that provides GPS tracking data. 
In context of this project, we propose usage of a simulation framework to test estimation methods of the current occupancy state of the parking lot. As it is a counterfactual process, where one parking decision influences the state of the whole system, A real-world use-case could hardly be used for validation. Therefore, we use SUMO, a microscopic traffic simulator to create repeatable traffic scenarios and test our approach under various conditions.

The paper is structured as follows. Section~2 describes our approach for parking lot occupancy estimation, afterwards we describe future directions of our study and discuss open issues in Section~3.

\FloatBarrier
\section{Estimation of Parking Lot Occupancy} 

\begin{figure}[tb]
	\begin{centering}
	\includegraphics[width=\textwidth]{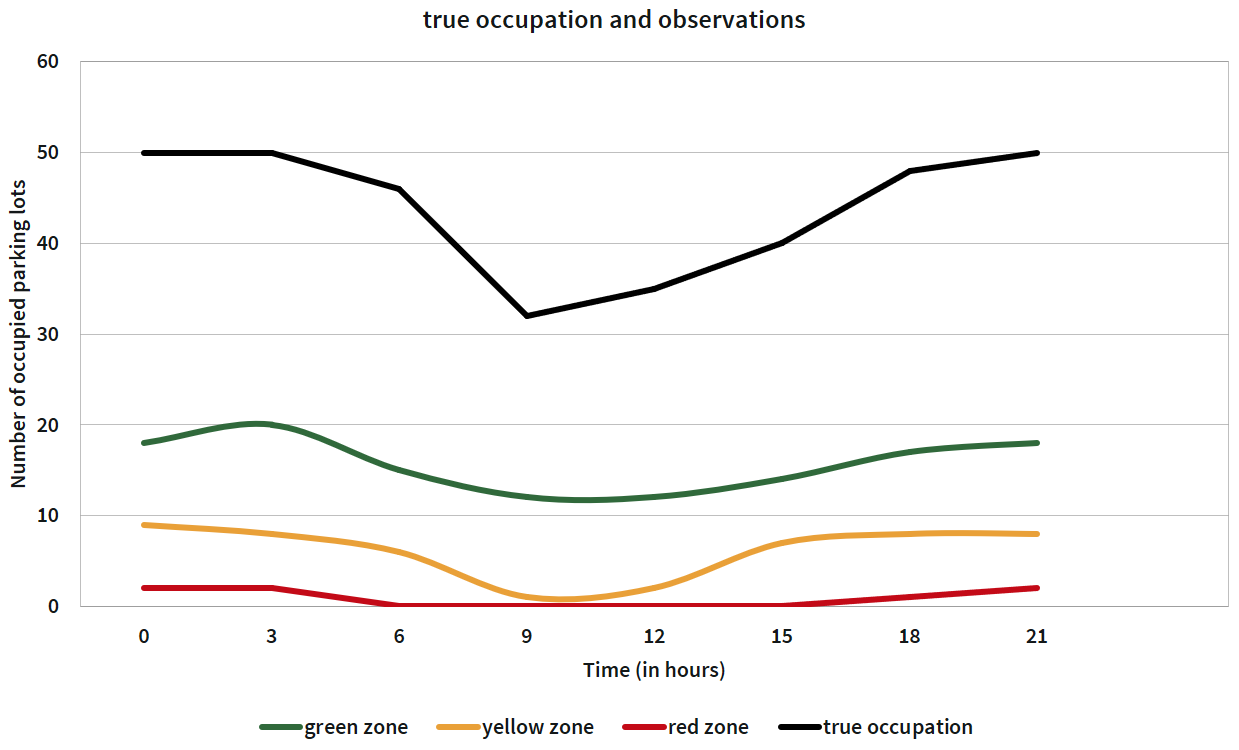}
	\caption{Observed occupancy versus true occupancy. Green zone: truck parking spaces. Yellow zone: spaces of the parking lot where trucks can park legally and which are not truck parking spaces. Red zone: parking spaces that are illegal for truck parking.}
	\label{fig:easychair-logo}
	\end{centering}
\end{figure}

\begin{figure}[tb]
	\begin{centering}
	\includegraphics[width=\textwidth]{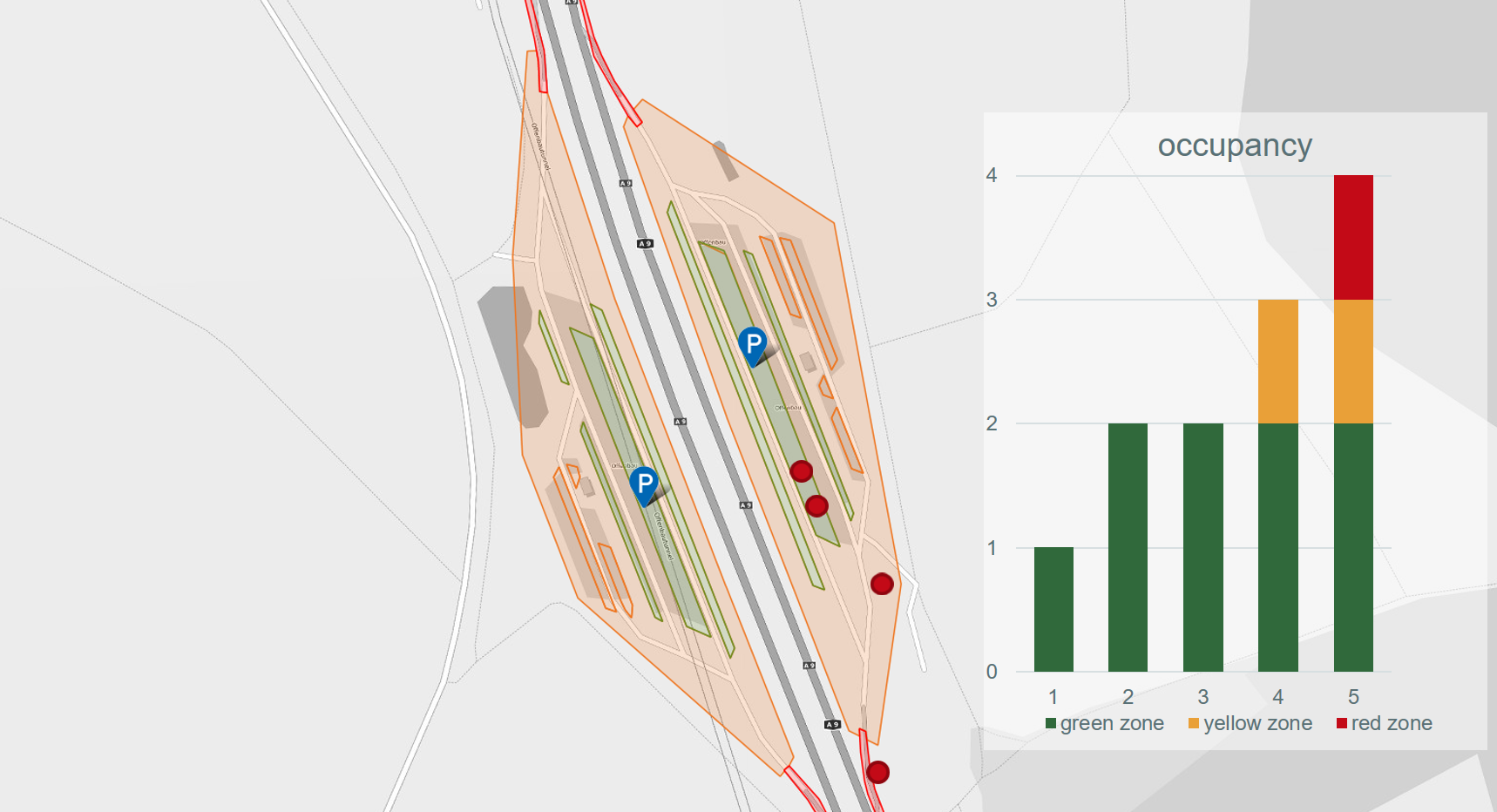}
	\caption{Offenbau resting lot with indicated zones. To the right an exemplified occupation of the different parking zones as observed from sensors.}
	\label{fig:offenbau}
	\end{centering}
\end{figure}

The prediction of available truck parking lots could benefit from the utilization of multiple data sources. The telemetry of the trucks (e.g., GPS traces) provides individual trajectories, and future demand for parking lots could be estimated. Also, the level of parking lot usage (eventually separated in zones) could be monitored and utilized for time series prediction of future parking lot availability (see Figure~\ref{fig:easychair-logo} and Figure~\ref{fig:offenbau}). 

However, in practice, these two data sources are very sparse as just a few trucks reveal their telemetry data. In contrast, our approach is the creation of a parking recommendation app. Its details are not focused on the paper at-hand, but here we present a method to estimate future parking lot availability, given such an app. The required data will be generated by a micro-simulation, which is adjusted to reflect the crucial features of the real-world. 
We study the estimation performance for various penetration rates. Since we are not interested in the actual number of free parking lots but need to identify the transitions of the parking lot levels, i.e., the change from an occupied one to a medium filled one to an empty one, it eases our problem. 
We, therefore, discretize the parking lot observations and assign three labels to it: \emph{full}, \emph{filled}, and \emph{empty}.




Recommending parking lots affects future traffic state and is thus hard to compare in a real-world scenario. To achieve repeatability of the experiments and compare different methods, we utilize a simulation and achieve reproducible scenarios. 
The simulation software SUMO, short for Simulation of Urban MObility, is an open source software package for the simulation of traffic flows~\cite{krajzewiczSumo2016}.
It is microscopic (simulating each individual vehicle), inter- and multimodal (supporting multiple types of vehicles and pedestrians), space-continuous and time-discrete.

In previous works, SUMO has been combined in other traffic related ressource allocation problems using  reinforcement learning. For example, \cite{roemer18} uses SUMO to train price policies of the smart electrical distribution grid for prevention of overloads due to electric vehicles. In \cite{liebig17b}, the authors use SUMO to compare various selfish routing regimes and propose usage of reinforcemnt learning for self-organization. 

We use SUMO to overcome poor data quality and get a repeatable traffic scenario where models can be compared and evaluated.
Without loss of generality, we model the traffic at the German highway A9 for the resting places G\"oggelsbuch
and Offenbau. The street network is obtained from OpenStreetMap and integrated in
SUMO, compare Figure~\ref{fig:offenbau}. Parking lots are manually modeled in this street network. 

To get a representative distribution of parking duration, we model it
by a mixed Gaussian and distinguish among short and long breaks calibrated to foating car data of trucks. Once a truck stops at a parking lot, a new route is assigned in the simulation, which will not park again.

To simulate sparse data availability by a recommendation app, we study two penetration rates of a potential recommendation app: 10 percent and 20 percent. In the following, we describe our procedure for the 10 percent case but 20 percent was performed similar. In the end we depict our results.

\begin{figure}[tb]
	\begin{centering}
	\includegraphics[width=\textwidth]{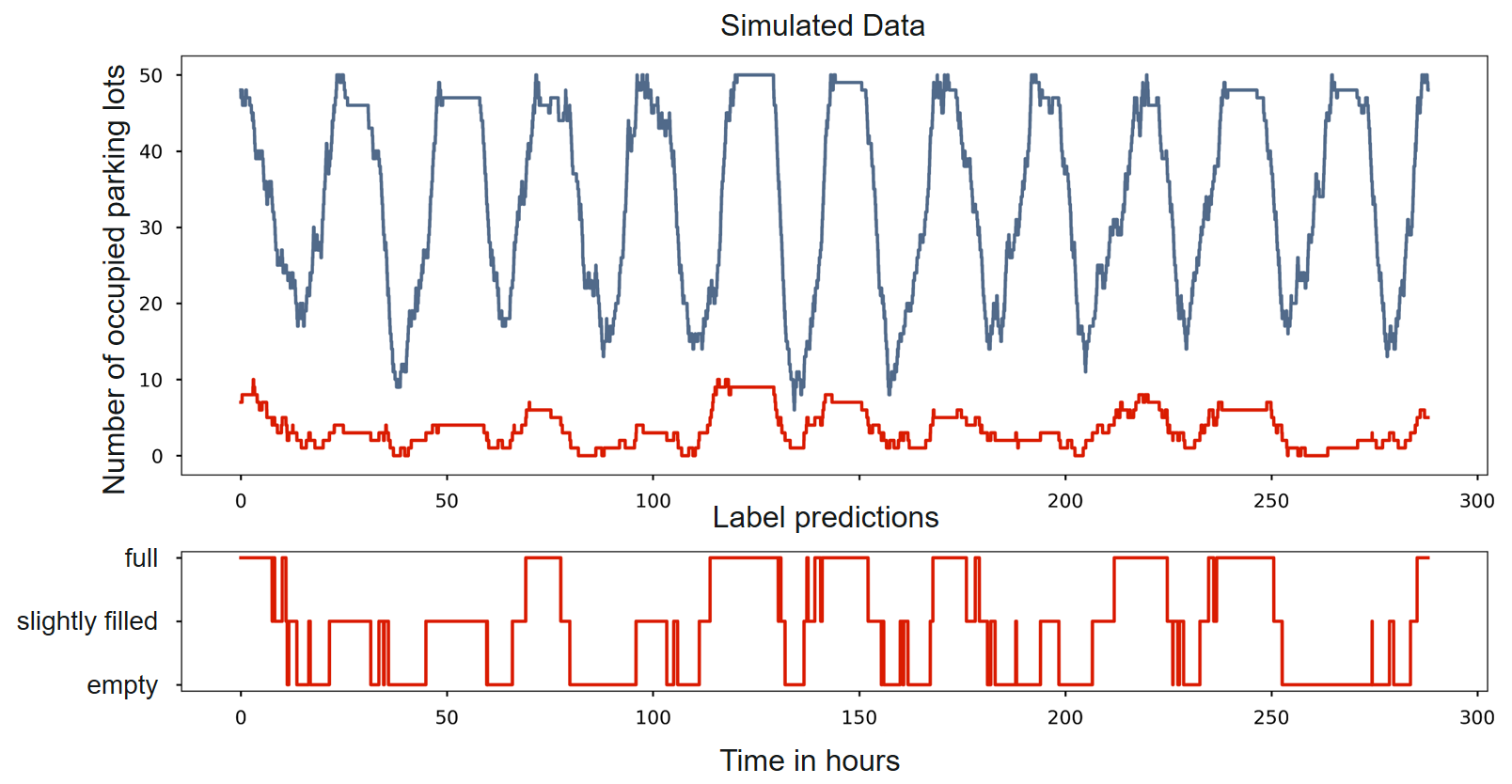}
	\caption{Simulated parking occupancy in blue and the sampled observations in red.}
	\label{fig:pred1}
	\end{centering}
\end{figure}

\begin{figure}[tb]
	\begin{centering}
	\includegraphics[width=\textwidth]{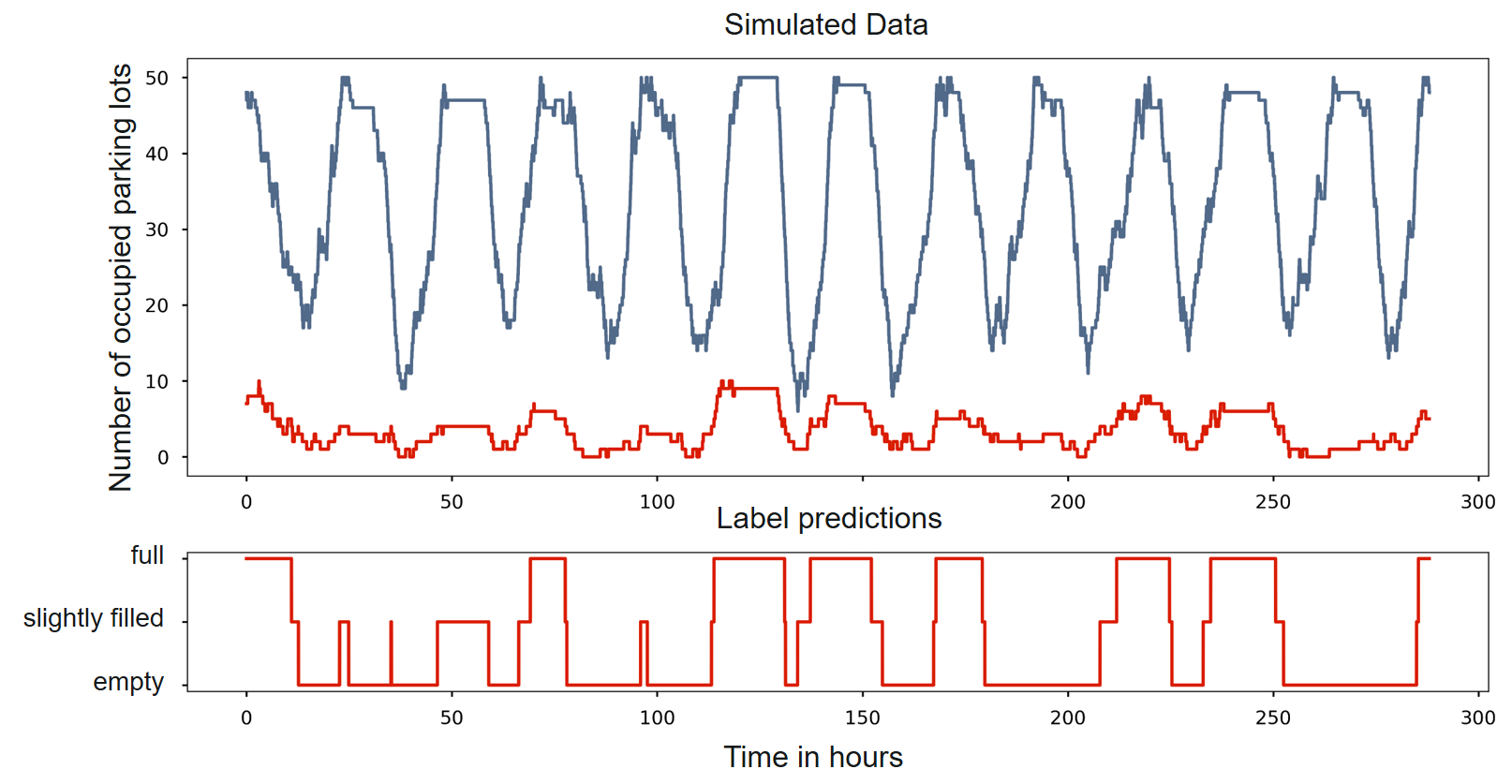}
	\caption{Estimation of Labels based on observed occupancy.}
	\label{fig:pred2}
	\end{centering}
\end{figure}

The simulatted time period is 22 days. The simulated parking lot occupancy is depicted in Figure~\ref{fig:pred1}, blue curve. In this figure, the red curve to the top indicates the simulated observations. Below are our estimates of the parking lot occupancy based on some threshold values. These threshold values are optimzed in a grid search and result in above 75 percent for slightly filled state, and above 95 percent for full state. 
The curve of the estimated labels are depicted below. Using some smoothing, we obtain the data as shown in Figure~\ref{fig:pred2}.

\begin{figure}[tb]
	\begin{centering}
	\includegraphics[width=\textwidth]{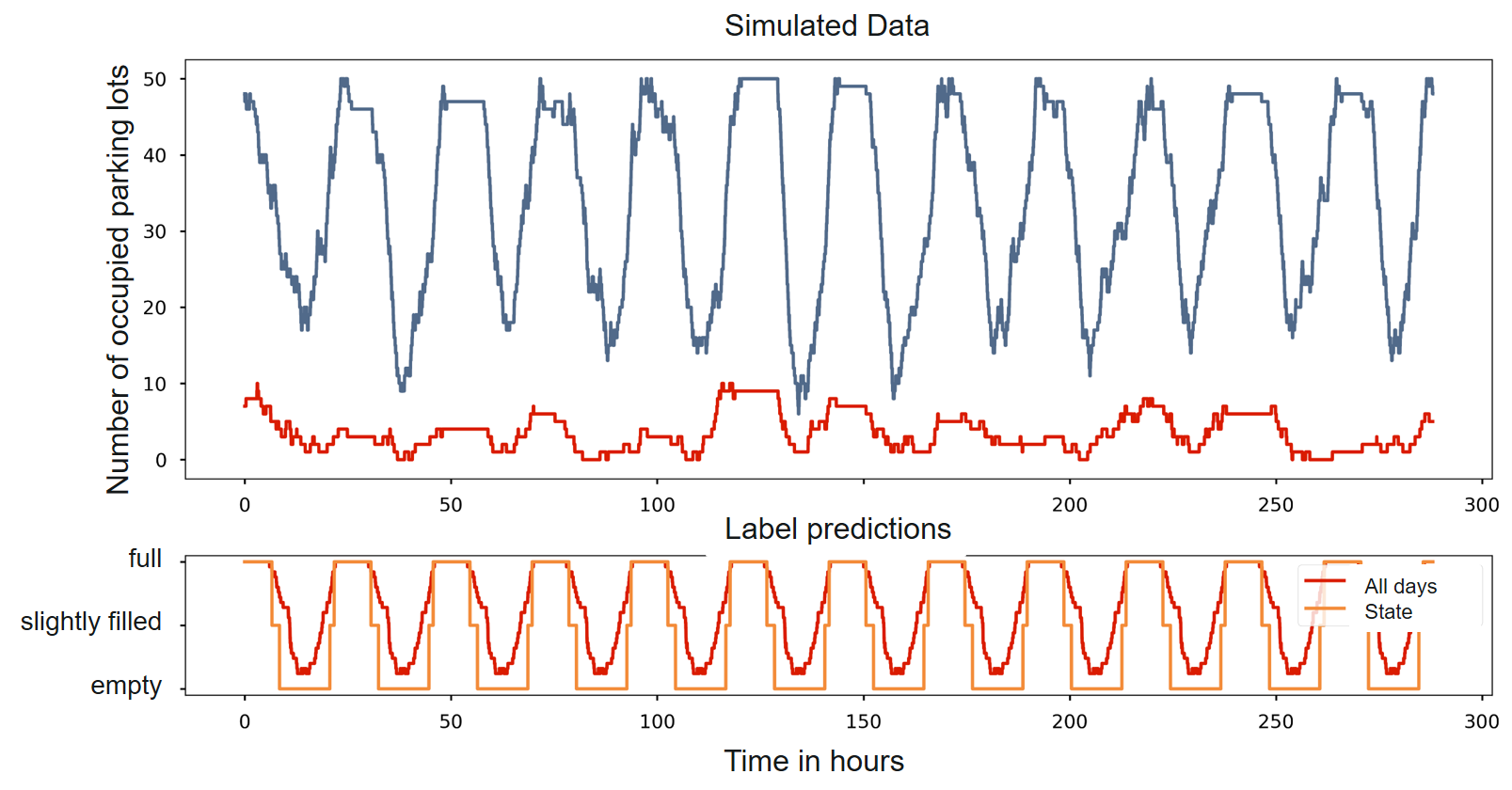}
	\caption{Estimation utilizing daily mean occupancy.}
	\label{fig:pred3}
	\end{centering}
\end{figure}

In order to achieve better estimates of current occupancy state, the mean of all days is computed (compare Figure~\ref{fig:pred3}) and occupancy state changes could be assigned more accurate assuming each day follows to some extent the mean day. We therefore, stretch the mean day to fit the current day best and compute again the occupancy states by using thresholds again (shown in Figure~\ref{fig:pred3} bottom, orange curve).

The estimation delay in minutes achieved by our method are depicted in Table~\ref{tab:ltbexample1} and Table~\ref{tab:ltbexample2}, bold are the values which are important for our application.
\begin{table}[htp]
	\begin{centering}
		\begin{tabular}{l|rrr}
		from \textbackslash to & full & slightly filled & empty \\ 
		\hline
		full & - & \textbf{81'} & - \\
		slightly filled & \textbf{63'} & - & 81' \\
		empty & - & 76' & - 
		\end{tabular}
		\caption{Delay of state estimation, with  penetration rate of 10\%.}
		\label{tab:ltbexample1}
	\end{centering}
\end{table}
\begin{table}[htp]
	\begin{centering}
		\begin{tabular}{l|rrr}
		from \textbackslash to & full & slightly filled & empty \\ 
		\hline
		full & - & \textbf{56'} & - \\
		slightly filled & \textbf{26'} & - & 120' \\
		empty & - & 30' & - 
		\end{tabular}
		\caption{Delay of state estimation, with  penetration rate of 20\%.}
		\label{tab:ltbexample2}
	\end{centering}
\end{table}


\FloatBarrier

\section{Summary and Future Work}
\label{sect:future-work}

In the paper-at-hand, we highlithed that SUMO perfectly fits a workflow to assess the parking lot occupancy estimation. We presented an approach for the estimation of free parking lot capacities based on sparse time-series observations. Our approach consists of the following steps: simulation, sampling of observed data, estimation of a daily mean curve, stretching and fitting this to particular days. We validate this approach in a simulation environment based on the microscopic traffic simulator SUMO. For two penetration rates, the utility of our approach is depicted at a particular resting lot. Analysis of more advanced estimation methods and easing the assumption that all days are similar (in real-world weekdays differ from weekends) are our next step. 

In future works, we extend the evaluation to other parking lots and incorporate Reinforcement Learning to prevent giving all drivers the same information and causing traffic jams based on our recommendation. 

\FloatBarrier

\section{Acknowledgments}
The authors were funded by the German Federal Ministry of Transport and Digital
Infrastructure (BMVI) program mFUND under grant number FKZ: 19F2037B ``Intelligent Truck Parking ITP''.

\label{sect:bib}
\bibliographystyle{plain}
\bibliography{paper}


\end{document}